\documentclass[prl,noshowpacs,english,superscriptaddress,twocolumn]{revtex4-1}
\usepackage{amsmath,amssymb,amsfonts,mathrsfs}
\usepackage{pgfplots,pgfplotstable,physics,epstopdf}
\usepackage{amsthm}
\usepackage{CJK}
\usepackage{bm}
\usepackage{babel}
\usepackage{graphicx}
\usetikzlibrary{shadings}
\usepackage[colorlinks=true,linkcolor=blue,anchorcolor=red,citecolor=blue,urlcolor=blue]{hyperref}
\usepackage{appendix}
\usepackage{tikz,pgf}
\usepackage{pdfpages}

\makeatletter
\AtBeginDocument{\let\LS@rot\@undefined}
\makeatother

\def \K {\hat{\mathcal{K}}}
\def \Z {\mathbb{Z}}
\def \H {\mathcal{H}}
\def \k {\bm{k}}

\def \T {\hat{T}}

\def \P {\hat{P}}

\begin{document}

\title{Boundary criticality of $PT$-invariant topology and second-order nodal-line semimetals}

\author{Kai Wang}
\email[These authors contributed  equally  to this work.]{}
\affiliation{National Laboratory of Solid State Microstructures and Department of Physics, Nanjing University, Nanjing 210093, China}

\author{Jia-Xiao Dai}
\email[These authors contributed  equally  to this work.]{}
\affiliation{National Laboratory of Solid State Microstructures and Department of Physics, Nanjing University, Nanjing 210093, China}

\author{L. B. Shao}
\email[]{lbshao@nju.edu.cn}
\affiliation{National Laboratory of Solid State Microstructures and Department of Physics, Nanjing University, Nanjing 210093, China}
\affiliation{Collaborative Innovation Center of Advanced Microstructures, Nanjing University, Nanjing 210093, China}

\author{Shengyuan A. Yang}
\address{Research Laboratory for Quantum Materials, Singapore University of Technology and Design, Singapore 487372, Singapore}

\author{Y. X. Zhao}
\email[]{zhaoyx@nju.edu.cn}
\affiliation{National Laboratory of Solid State Microstructures and Department of Physics, Nanjing University, Nanjing 210093, China}
\affiliation{Collaborative Innovation Center of Advanced Microstructures, Nanjing University, Nanjing 210093, China}

\begin{abstract}
For conventional topological phases, the boundary gapless modes are determined by bulk topological invariants. Based on developing an analytic method to solve higher-order boundary modes, we present $PT$-invariant $2$D topological insulators and $3$D topological semimetals that go beyond this bulk-boundary correspondence framework. With unchanged bulk topological invariant, their first-order boundaries undergo transitions separating different phases with second-order-boundary zero-modes. For the $2$D topological insulator, the helical edge modes appear at the transition point for two second-order topological insulator phases with diagonal and off-diagonal corner zero-modes, respectively. Accordingly, for the $3$D topological semimetal, the criticality corresponds to surface helical Fermi arcs of a Dirac semimetal phase. Interestingly, we find that the $3$D system generically belongs to a novel second-order nodal-line semimetal phase, possessing gapped surfaces but a pair of diagonal or off-diagonal hinge Fermi arcs.
\end{abstract}


\maketitle

{\color{blue}\textit{Introduction}.} 
Topological phases have been one of the most actively expanding fields in physics during the last fifteen years, including both fully gapped topological systems~\cite{Volovik:book,Kane-RMP,XLQi-RMP} such as topological insulators (TIs), and gapless systems~\cite{Vishwanath-RMP} such as Weyl and Dirac semimetals. The classification of topological phases crucially depends on symmetry.
As basic results in the field, topological phases with fundamental symmetries including time reversal $T$ and {charge conjugation} $C$ have been completely classified for both gapped~\cite{Schnyder2008,Kitaev2009AIP} and gapless~\cite{HoravaPRL05,ZhaoYXWang13prl,ZhaoYXWang14Septprb} systems in the framework provided by the real $K$-theory~\cite{Atiyah-KR} and the tenfold Altland-Zirnbauer symmetry classes~\cite{ShinseiRyu-RMP}.

Recently, the classification has been extended to topological phases protected by combined symmetries $PT$ and $CP$ with $P$ the spatial inversion, by using the orthogonal $K$-theory~\cite{ZhaoWang16Aprprl,Karoubibook,Atiyah-KR}. This constitutes an important starting point, because these symmetries are fundamental to a large class of interesting systems, not only solid quantum materials~\cite{ZhaoLu17Aprprl,Tomas-PRB,Band-Combinatorics,B-J-Yang19APRPRX,ZhaoYang19prl,Non-Abelain-NL,Wangzhijun2019prl}, but also photonic, cold-atom, classical acoustic and circuit systems~\cite{Topo_ColdAtom_Review,Yu_Zhao_NSR,YangZJ2015prl,Ronny_2018np,Serra_Garcia_2018nature,Peterson_2018nature,Ronny_2019arxiv,Ozawa2019RMP,MaGC_2019nature}, where the physics of $PT$ symmetry is under active research.
Now, an important question is whether the topology governed by these symmetries will manifest any unique feature distinct from all known examples before.

In this Letter, we uncover one such distinct feature. This concerns the perhaps most central property of topological phases --- the bulk-boundary correspondence, which is usually understood as: the bulk topological invariant completely determines the boundary topological modes. For conventional ($T$-invariant) TIs, this correspondence has been rigorously established, and described by an index theorem \cite{ZhaoYXWang14Septprb,ZhaoYXWang13prl,ShinseiRyu-RMP}. Here, we show that this common belief no longer holds for $PT$-invariant topological systems. Specifically, we demonstrate that the bulk invariant of a $PT$-invariant real system \emph{cannot} uniquely determine the form of the boundary topological modes, rather, it determines \emph{a boundary criticality}, {meaning that there are multiple phases with distinct boundary modes, and the critical states between them correspond to topological phase transitions at the boundary.} In 2D, it corresponds to an edge criticality with gapless helical edge modes separating two second-order TI phases \cite{Mele2013prl,TaylorHughes2017Science,Songzida2017prl,Brouwer2017prl,Schindler2018np} with corner zero-modes. In 3D, it gives a surface criticality with helical surface Fermi arcs separating two topological semimetal phases with hinge Fermi arcs. Interestingly, the 3D semimetal phase possesses bulk nodal loops and a single pair of $PT$-related hinge Fermi arcs, representing a novel second-order semimetal phase not known before. An analytical approach for solving higher-order boundary modes at uneven boundaries is also developed in this work.

{\color{blue}\textit{Edge criticality of 2D PT-invariant real Chern insulator}.}
The fundamental symmetry considered in this work is the combined symmetry $PT$ (while the individual $P$ and $T$ may be violated), where $T$ is for systems without spin-orbit coupling. In momentum space,
the symmetry is represented by $ \P\T=U\K$, with $U$ an unitary operator and $\K$ the complex conjugation, satisfying $(\P\T)^2=1$. Importantly, $PT$ preserves the momentum $\bm{k}$. Without loss of generality, we choose the representation $\hat{P}\hat{T}=\K$ below~\cite{PT-symmetry}. 

This $PT$ symmetry imposes an reality condition on the system, i.e., the $PT$-invariant Hamiltonian in momentum space must be a real matrix for each $\k$. To have a nontrivial topology under this condition, it is convenient to consider a four-band model expressed by the real Dirac matrices~\cite{ZhaoWang16Aprprl}. For $4\times 4$ Dirac matrices $\gamma^{\mu}$ satisfying $\{\gamma^\mu,\gamma^\nu\}=2\delta^{\mu\nu}$  with $\mu,\nu=1,2,\cdots,5$, at most three of them, say $\gamma^{1,2,3}$, can be real, while the other two, $\gamma^{4,5}$, are then purely imaginary. For instance, we may choose $\gamma^1=\sigma_0\otimes\tau_3$, $\gamma^2=\sigma_2\otimes\tau_2$, $\gamma^3=\sigma_0\otimes\tau_1$, and $\gamma^{4,5}=\sigma_{1,3}\otimes\tau_2$, with $\sigma_i$ and $\tau_i$ two sets of Pauli matrices and $\sigma_0$ the identity matrix.

Let's consider the following 2D $PT$-invariant Hamiltonian,
\begin{equation}\label{H0}
\H_0(\k)=\sin k_x\gamma^1+\sin k_y\gamma^2+(M-\cos k_x-\cos k_y)\gamma^3,
\end{equation}
defined on a square lattice. An interesting observation is that the Hamiltonian (\ref{H0}) has exactly the same form as the model for a 2D $T$-invariant TI~\cite{QiPRB2008}, where it was shown that the model is nontrivial and hosts gapless edge helical states when $M\in (-2,0)\cup(0,2)$. However,
there is a fundamental distinction between the two contexts: when interpreted as a $T$-invariant TI~\cite{QiPRB2008}, the defining symmetry is $T$ with $\hat{T}^2=-1$; but here, the defining symmetry is $PT$,
so our model \eqref{H0} must be given a completely different interpretation.

The interpretation can be inferred from the topological invariant enabled by the respective symmetry. While $T$ protects the $\Z_2$ invariant for the $T$-invariant TI, the bulk invariant protected by $PT$ is the real Chern number formulated in Ref.~\cite{ZhaoLu17Aprprl}: \begin{equation}\label{R-Chern}
\nu_R=\frac{1}{4\pi}\int d^2k~\mathrm{tr}(I\mathcal{F}_R)\mod 2.
\end{equation}
Here, the Berry curvature $\mathcal{F}_R=(\nabla_{\bm k}\times\bm{\mathcal{A}})_{z}$, where $\bm{\mathcal{A}}_{\alpha\beta}=\langle \alpha,\k|\nabla_{\bm k}|\beta,\k\rangle$ is the \emph{real} Berry connection derived from the valence band eigenstates $|\alpha,\k\rangle$ ($|\beta,\k\rangle$), which preserve the $PT$ symmetry, i.e., $|\alpha,\k\rangle^*=|\alpha,\k\rangle$~\cite{Real-Chern}. $I=-i\sigma_2$ is the generator of the $SO(2)$ rotation in the $2$D Euclidean space spanned by the real valence eigenstates~\cite{Pauli-Matrices}. For $M\in (-2,0)\cup(0,2)$, we have a nontrivial $\nu_R=1$, so the resulting state should be interpreted as a 2D $PT$-invariant real Chern insulator (RCI).

\begin{figure}
	\includegraphics[scale=0.3]{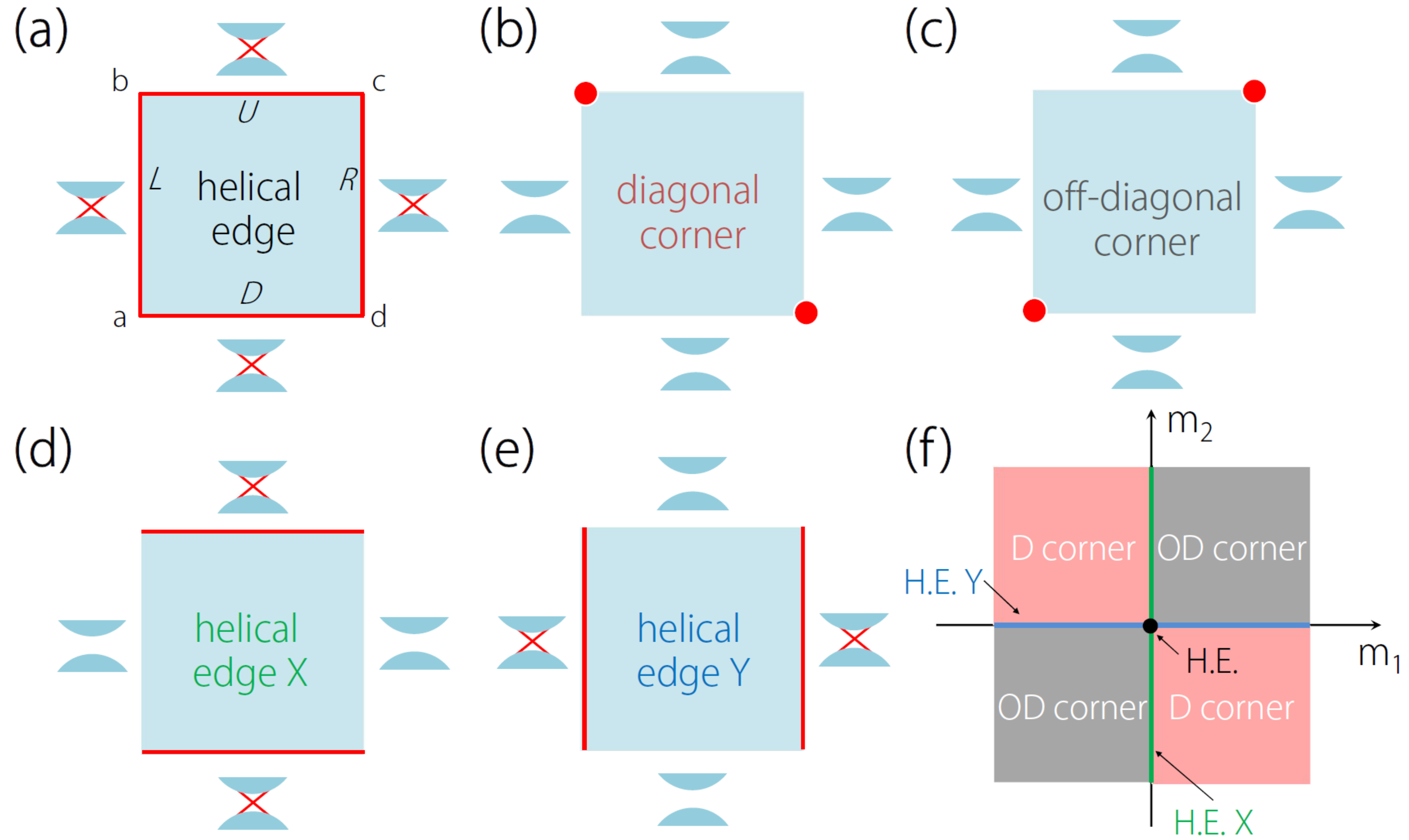}
	\caption{ (a-e) illustrate possible topological boundary-mode configurations for the 2D $PT$-invariant RCI. (a) The critical state [Eq.~(\ref{H0})] has helical edge modes over all edges. (b) and (c) illustrate the two second-order TI phases with a single pair of corner zero-modes on diagonal and off-diagonal corners, respectively. (d) and (e) are states at phase boundaries between (b) and (c), which has helical edge modes only on a single pair of edges. (f) shows the phase diagram with respect to $m_1$ and $m_2$, where ``D", ``OD", and ``H.E." stand for diagonal, off-diagonal, and helical edge, respectively.
	}\label{2D_Insulator}
\end{figure}

For a 2D $T$-invariant TI, the bulk $\Z_2$ invariant dictates the existence of gapless edge helical modes. Now, the question is: Will the same bulk-boundary correspondence work for the $PT$-invariant RCI? We find the answer is negative. Instead, the nontrivial bulk invariant $\nu_R$ determines a phase transition between two second-order TI phases. More specifically, there exist $PT$-invariant perturbations that can gap the edge helical modes of \eqref{H0} and lead to distinguishable phases of second-order TIs with corner zero-modes [with the edge helical states representing the corresponding edge critical point, see Fig.~\ref{2D_Insulator}(a) and (f)]. It is emphasized that in the whole process of the edge phase transition, the bulk gap is always open, therefore the bulk invariant $\nu_R$ is unchanged.

One may infer a possibly gapped edge directly from the form of Eqs.~\eqref{H0} and \eqref{R-Chern}. By the {unitary transformation $e^{-\frac{\pi}{4}\gamma^2\gamma^5}$}, Eq.~\eqref{H0} is diagonalized into two blocks of (conventional) Chern insulators with Chern numbers $\pm 1$, respectively. Meanwhile, $\mathcal{F}_R$ and $I$ are also diagonalized, corresponding to the two Chern insulators. But there exist perturbations, which respect $PT$ (so are symmetry-allowed) but lead to off-diagonal terms, and therefore can gap the edge helical modes with the real Chern number $\nu_R$ being preserved.

Particularly, let's consider adding to Eq.~\eqref{H0} the following perturbations
\begin{equation}\label{perturbations}
\Delta\H=i\gamma^1(m_1\gamma^4+m_3\gamma^5)+i\gamma^2(m_2\gamma^4+m_4\gamma^5),
\end{equation}
which are the only $PT$-invariant \emph{relevant} quadratic perturbations for \eqref{H0}.  Among the totally ten $PT$-invariant terms, the other two nontrivial ones are given by $i\gamma^3\gamma^{4,5}$. However, both of them commute with the kinetic terms of \eqref{H0}, so they only modify the mass term in \eqref{H0} and are irrelevant here. In contrast, each term in Eq.~\eqref{perturbations} anti-commutes with one of the kinetic terms, and therefore affects the edge helical critical point.

Based on the analytical method that we shall present, $\Delta \mathcal{H}$ generally drives the edge critical point into second-order TI phases with localized corner modes [see Fig.~\ref{2D_Insulator}(b,c)]. If further requiring the corner modes to be at zero energy, one can show that this occurs if
and only if the following concise equation holds,
\begin{equation}\label{zm-condition1}
\left(m_1,m_3\right)=\alpha\left(m_2,m_4\right)
\end{equation}
with $\alpha\ne 0$ and at least one of the $m_i$'s is nonzero. The two second-order TI phases separated by the edge critical point are distinguished by the location of the corner zero-modes. Considering a square-shaped sample respecting the $PT$ symmetry as in Fig.~\ref{2D_Insulator}, for the case with  $\alpha>0$ ($\alpha<0$), the corner zero-modes are found at corners $a, c$ ($b, d$) but not at $b, d$ ($a, c$), as shown in Fig.~\ref{2D_Insulator}(c) (Fig.~\ref{2D_Insulator}(b)), which we refer to as the off-diagonal (diagonal) second-order TI.
This pair of corner modes are connected by $PT$, so they remain degenerate, even if their energy may deviate from zero for the most general case beyond (\ref{zm-condition1}). It is also worth noting that when Eq.~\eqref{zm-condition1} holds, the whole Hamiltonian $\mathcal{H}=\mathcal{H}_0+\Delta\mathcal{H}$ anti-commutes with $(m_1\gamma^4+m_3\gamma^5)$, which represents an emergent chiral symmetry setting the mid-gap modes exactly at zero energy.

Figure~\ref{2D_Insulator}(f) shows the phase diagram with respect to $m_1$ and $m_2$ (with $m_3=m_4=0$). Interestingly, the phase boundary with $m_1=0$ ($m_2=0$) corresponds to a crystalline TI, with gapless helical edge modes only for the $x$-edges ($y$-edges), and with gapped spectrum for the $y$-edges ($x$-edges), as illustrated in Fig.~\ref{2D_Insulator}(d) (Fig.~\ref{2D_Insulator}(e)). This can be intuitively understood from the physically meaning of the partial mass terms in \eqref{perturbations}, as $m_{1,3}$ ($m_{2,4}$) is the mass term only for the $x$-direction ($y$-direction) kinetic term in Eq.~\eqref{H0}~\cite{Supp,LiuTaoPRL2019}. Again, we stress that for the whole phase diagram, the bulk gap is not closed and hence $\nu_R=1$ remains unchanged.

The above discussion confirms that as a topological state, the $PT$-invariant RCI does not possess the usual bulk-boundary correspondence. Namely, the bulk invariant $\nu_R$ cannot uniquely determine the boundary modes, but dictates an edge criticality. This feature distinguishes the system from conventional TIs.

\begin{figure}
	\centering
	\includegraphics[width=0.49\textwidth]{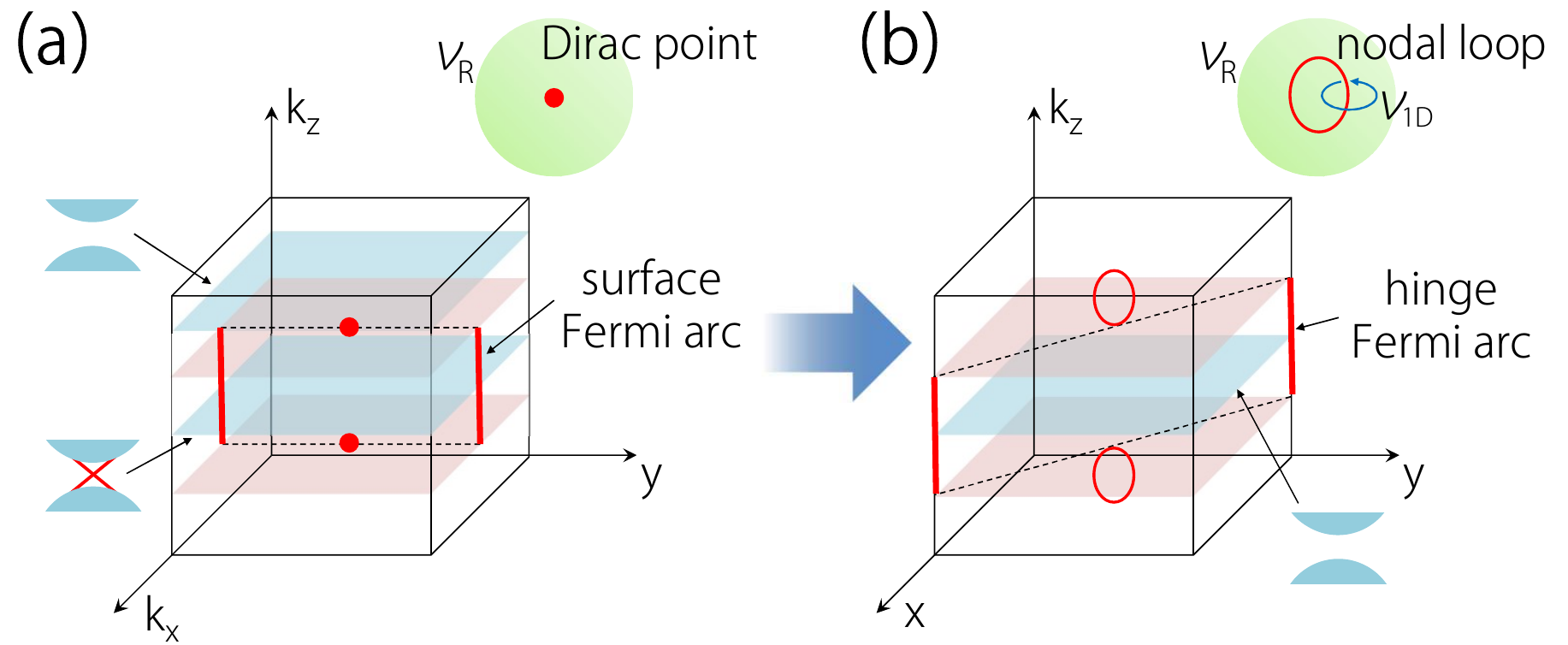}
	\caption{(a) The critical state of (\ref{3D}) has two bulk Dirac points and surface helical Fermi arcs on the side surfaces. Spectra for representative 2D subsystems are illustrated. The inset on the top-right shows each Dirac point carries a nontrivial $\nu_R$ defined on a sphere surrounding it. (b) The generic phase of the system is a second-order nodal-line semimetal with a single pair of hinge Fermi arcs. The inset shows each Dirac point evolves into a nodal loop, which retains the nontrivial $\nu_R$ and meanwhile acquires the 1D topological charge $\nu_\text{1D}$. }
	\label{semimetal}
\end{figure}

{\color{blue} \textit{Second-order nodal-line semimetal}.} Our theory can be generalized to 3D, with even richer physical consequences. Particularly, we will show the surface criticality leads to a second-order nodal-line semimetal phase hosting a single pair of hinge Fermi arcs [see Fig.~\ref{semimetal}(b)], distinct from all previously known examples~\cite{Linmao2018prb,Ezawa2018prb,wieder2020strong} which are of nodal-point type.

The easiest extension of the 2D model \eqref{H0} to 3D is given by
\begin{equation}\label{3D}
\H^\text{3D}_0=\sin k_x\gamma^1+\sin k_y\gamma^2+(M-\sum_{i=1}^3\cos k_i)\gamma^3.
\end{equation}
Physically, it can be realized by a layer construction with layers of $2$D  RCIs and trivial insulators stacked in an alternating manner. The similar approach has been used to construct Weyl and Dirac semimetals before~\cite{WSTImultilayer}. Provided that the trivial insulator has a much larger gap than the RCI, the low-energy physics will correspond to the RCI layers with the tunnelings between them through the trivial insulator layers. We assume that each RCI is a two-layer system with each layer consisting of two sublattices. With $\sigma$'s acting on the sublattice space and $\tau$'s operating on the layer space, the interlayer hopping term is simply $(t\psi_{\bm{k}_\perp,j+1}^\dagger\sigma_0\otimes\tau_1\psi_{\bm{k}_\perp,j}+\mathrm{h.c.})$, where $\bm{k}_\perp=(k_x,k_y)$ and $j$ labels the layers. Recalling $\gamma^3=\sigma_0\otimes\tau_1$, the interlayer hopping gives exactly the $\cos k_z \gamma^3$ term in Eq.~\eqref{3D}.


$\mathcal{H}^\text{3D}_0$ alone describes a $3$D real Dirac semimetal when $1<M<3$, {where there are two real Dirac points residing at $k_z=\pm K_z=\pm \arccos(M-2)$ on the $k_z$-axis} [Fig.~\ref{semimetal}(a)]. The $k\cdot p$ models for the two Dirac points are given by
\begin{equation}\label{Dirac-points}
\mathcal{H}^{\pm}_0(\bm{q})=q_x\gamma^1+q_y\gamma^2\pm v_z q_z\gamma^3,
\end{equation}
where $\bm{q}$ is measured from each Dirac point, and $v_z=\sin K_z$.
Each point carries a nontrivial real Chern number $\nu_R$, defined on a sphere surrounding it. As discussed in Ref.~\cite{ZhaoLu17Aprprl}, the topological charges of the two Dirac points lead to surface helical Fermi arcs confined by $k_z\in(-K_z,K_z)$ on the side surfaces, e.g., the $x$-$z$ and $y$-$z$ surfaces for a cubic sample [Fig.~\ref{semimetal}(a)]. These surface helical modes can also be readily understood from our construction. For two $2$D $k_x$-$k_y$ subsystems on two sides of a chosen Dirac point, one of them must be topological nontrivial (as a RCI) and the other trivial, since the difference of their $\nu_R$ is equal to the nontrivial Chern number of the Dirac point. Particularly, in Eq.~\eqref{3D}, for each fixed $k_z\in(-K_z,K_z)$, the $2$D subsystem $\mathcal{H}^\text{3D}_0(\bm{k}_\perp, k_z)$ is just a $2$D RCI of Eq.~\eqref{H0}, with helical edge modes. It is these edge modes that trace out the Fermi arcs on the side surfaces of the 3D system.

From this analysis, it also becomes clear that
$\mathcal{H}^\text{3D}_0$ must correspond to a surface criticality. That is, the surface helical modes are not stable:
If the relevant perturbations in $\Delta\mathcal{H}$ [Eq.~\eqref{perturbations}] are turned on, the surface helical Fermi arcs will generically transform into off-diagonal or diagonal hinge Fermi arcs as illustrated in Fig.~\ref{semimetal}(b). And corresponding to the phase diagram in Fig.~\ref{2D_Insulator}(f), the phases with off-diagonal and diagonal hinge Fermi arcs are separated by anisotropic critical states with the surface helical Fermi arcs only existing on the $x$-$z$ or $y$-$z$ surfaces.

The perturbations $\Delta\mathcal{H}$ simultaneously affect the bulk energy spectrum. The Dirac points are perturbed as $\mathcal{H}^{\pm}=\mathcal{H}_0^{\pm}+\Delta\mathcal{H}$, and one finds that each Dirac point is spread into a nodal loop parallel to the $k_z$-axis [Fig.~\ref{semimetal}(b)], as both terms in $\Delta\mathcal{H}$ commutes with $\gamma^3$ while each anti-commutes with $\gamma^1$ or $\gamma^2$. Thus, the generic phase of our system is a $PT$-invariant second-order nodal-line semimetal with a single pair of hinge Fermi arcs.

Interestingly,
each nodal loop here carries two topological charges. First, because they originate from the Dirac points, they inherit the $2$D topological charges $\nu_R$ of the Dirac points (defined on a sphere enclosing each loop). Second, similar to conventional nodal lines, they have the $1$D topological charge 
defined by the Berry phase $\nu_\text{1D}$
on a small circle $S^1$ transversely surrounding them [inset of Fig.~\ref{semimetal}(b)]. The $1$D topological charges lead to the usual drumhead surface states bounded by the projections of the loops in the surface Brillouin zone \cite{YangSY2014prl,WengHM2015prb}.
Accordingly, the aforementioned second-order hinge Fermi arcs are in turn bounded by the drumhead states. Nevertheless, it is noteworthy that the drumhead states are extensive in real space, whereas the hinge Fermi arcs appear only on diagonal or off-diagonal hinges.



{\color{blue}\textit{Analytic method for solving corner zero-modes}.} As promised in the discussion of RCI, we now present an analytic method for solving the corner zero-modes.

The key point of the method is the following. The nontrivial $\nu_R$ implies for each edge the effective theory is a $2\times 2$ massive Dirac model, for instance, $\mathcal{H}_\text{edge}^\text{eff}=k_x\sigma_1+m\sigma_2$. But when considering a corner, it is not clear how to formulate a proper boundary condition for the two edges joined at the corner, because the $\sigma$ matrices for the two edges generally operate on different bases. Thus,
for sharp corners, we have to develop a method with the microscopic information, which distinguishes our method from previous ones based solely on edge effective theory.

For concreteness, considering the square geometry in Fig.~\ref{2D_Insulator}, {the effective Hamiltonians $\H^{\xi}$ of the four edges can be obtained by a projection from the bulk Hamiltonian $\mathcal{H}_\text{bulk}=\mathcal{H}_0+\Delta H$:
\begin{equation}\label{EFF2}
\H^{\xi}=\Pi^{\xi}\H_\text{bulk}\Pi^{\xi},
\end{equation}
where $\xi=U,D,R,L$ labels the four edges, and the projectors are}
\begin{equation}\label{projector}
\Pi^{U/D}=\frac{1}{2}(1\pm i\gamma^2\gamma^3),\quad \Pi^{R/L}=\frac{1}{2}(1\pm i\gamma^1\gamma^3),
\end{equation}
{satisfying $ \left(\Pi^{\xi}\right)^2=\Pi^{\xi} $.} 
The details for derivng these projectors can be found in the Supplemental Material (SM)~\cite{Supp}.
These projectors are related by $ PT $ as $ (\hat{P}\hat{T}) \Pi^{L,D} (\hat{P}\hat{T})^{-1} =\Pi^{R,U} $.

Now consider the corner $a$ ($c$ is related to $a$ by $PT$).
From Eq.~\eqref{EFF2}, the effective Hamiltonians for the two relevant edges are given by
\begin{equation}\label{EFF3}
\begin{split}
\H^D&=\gamma^1(\sin k_x+im_1\gamma^4+im_3\gamma^5)\Pi^D,\\
\H^L&=\gamma^2(\sin k_y+im_2\gamma^4+im_4\gamma^5)\Pi^L,
\end{split}
\end{equation}
where $k_{x,y}\in [-\pi,\pi)$ correspond to edge modes if and only if $|M-\cos k_{x,y}|<1$. For technical convenience, let's focus on the parameter region where $0<M<2$ and $2-M$ is sufficiently small but still much greater than $m$'s. Then, the edge effective Hamiltonians can be simplified by replacing $\sin k_{x,y}$ with $k_{x,y}$ or $-i\partial_{x,y}$.
Since the spectrum of the boundary Hamiltonian is gapped, if there exists a zero-mode at $ a $, the state must be localized and  decay exponentially away from the corner. Therefore, we can adopt the ansatzs for the zero-mode along the two edges as
\begin{equation}\label{zm}
\psi^{D}(x)=\psi_0^{D} e^{-\lambda_xx},\quad \psi^{L}(y)=\psi_0^{L} e^{-\lambda_y y},
\end{equation}
where the decay rates $ \lambda_{x,y}>0 $. Then, the corner zero-mode can be solved by the equations:
\begin{equation}\label{zm-1}
\begin{split}
(m_1\gamma^4+m_3\gamma^5)\psi_0^D&=-\lambda_x\psi_0^D,\\
(m_2\gamma^4+m_4\gamma^5)\psi_0^L&=-\lambda_y\psi_0^L,
\end{split}
\end{equation}
with the boundary condition $\psi_0^D=\psi_0^L$ that connects $\psi^{L/D}$ at corner $a$.

The decay rates are obtained as $ \lambda_x=\sqrt{m_1^2+m_3^2} $ and $ \lambda_y=\sqrt{m_2^2+m_4^2} $. As a result, $ \psi_0 $ is simultaneously the eigenstate with eigenvalue $ -1 $ for operators $ \Lambda_1= ( m_1\gamma^4+m_3\gamma^5) /\lambda_x $ and $ \Lambda_2=( m_2\gamma^4+m_4\gamma^5) /\lambda_y $. 
By the anti-commutation relations of $\gamma$ matrices,  $ \Lambda_1 $ and $\Lambda_2$ share the same set of eigenstates $ \{\psi_0, \gamma^1\psi_0,\gamma^2\psi_0,\gamma^1\gamma^2\psi_0\} $, so they must commute with each other, resulting in $m_1m_4=m_2m_3$ or equivalently Eq.~\eqref{zm-condition1}. Substituting Eq.~\eqref{zm-condition1} into $\Lambda_1$ and using again the fact that $\Lambda_1$ and $\Lambda_2$ have the eigenstate $\psi_0$ with the same eigenvalue $-1$, we find that $\alpha=\lambda_x/\lambda_y>0$. Hence, we arrive at the conclusion that once Eq.~\eqref{zm-condition1} holds with $\alpha>0$ and $(m_1,m_3)\ne 0$, there are $PT$-related corner zero-modes at $a$ and $c$, corresponding to the off-diagonal second-order TI phase.
By a parallel argument, the conditions for the diagonal phase can also be obtained. 

We proceed to interpolate the two second-order TI phases to justify the phase diagram in Fig.~\ref{2D_Insulator}(f). From analytic solutions in Eq.~\eqref{zm}, we observe that in the course of adiabatically turning down $\alpha$ [with fixed $(m_2,m_4)$], the distribution of the zero-mode at corner $a$ ($c$) tends to be more and more spread over edge $D$ $(U)$. Meanwhile, the distribution over $L$ $(R)$ is unchanged. In the limit of $\alpha\rightarrow 0$, $\psi^D(x)$ in Eq.~\eqref{zm} is no long a corner state. Hence, the system approaches the crystalline TI state at the phase boundary, with a single pair of helical $x$-edges [Fig.~\ref{2D_Insulator}(d)].

{\color{blue}\textit{Discussion}.} For
the tenfold symmetry classification, there is an elegant mathematical framework accounting for the faithful bulk-boundary correspondence, namely the Teoplitz index theorem and $K$-theory~\cite{Volovik:book,Prodan-TI,Kitaev2009AIP}. However, as we show in this work, topological insulators related to spatial symmetries, for instance $PT$ symmetry here,
lie beyond this framework,
and result in a much richer boundary physics.  

The study is closely related to practical systems. {The physics not only applies to solid materials with $PT$ symmetry and negligible spin-orbit coupling (such as carbon allotropes), but also to $PT$-invariant bosonic and classical systems~\cite{Topo_ColdAtom_Review,Yu_Zhao_NSR,YangZJ2015prl,Ronny_2018np,Ronny_2019arxiv,
Serra_Garcia_2018nature,Peterson_2018nature,Ozawa2019RMP,MaGC_2019nature}.}
In addition, we briefly discuss typical $PT$-invariant non-Hermitian perturbations in the SM~\cite{Supp}.

{Finally, similar to previous analysis on second-order TI phases~\cite{Songzida2017prl,Brouwer2017prl}, the topological states here are robust against perturbations from boundary defects or irregularities,
as long as the bulk and edge gaps are not fully closed.}



\begin{acknowledgements}
  The authors acknowledge the support from the National Natural Science Foundation of China under Grant (No.11874201 and No.11704180), the Fundamental Research Funds for the Central Universities (No.14380119) and the Singapore Ministry of Education AcRF Tier 2 (MOE2019-T2-1-001).
\end{acknowledgements}



\bibliography{ref-PT-RTIs-RDSMs}

\begin{thebibliography}{47}
\bibitem[{PT-()}]{PT-symmetry}
\bibinfo{note}{Once $(\hat{P}\hat{T})^2=1$ is satisfied, there always exists a
  unitary transformation $U$, such that
  $UU_{PT}\hat{\mathcal{K}}U^\dagger=e^{i\phi}\hat{\mathcal{K}}$. But
  $e^{i\phi}\hat{\mathcal{K}}$ is equivalent to $\hat{\mathcal{K}}$, since they
  impose the same constraint to the Hamiltonian.}
\bibitem[{Rea()}]{Real-Chern}
\bibinfo{note}{The topological nontrivialness here means the inability to have
  real valence wavefunctions globally defined over the Brillouin zone while
  maintaining the periodicity. The transition functions for coordinate charts
  are characterized by $\pi_1(SO(N))$, where $N$ is the number of valence
  bands. Since $\pi_1(SO(2))=\mathbb{Z}$ and $\pi_1(SO(N))=\mathbb{Z}_2$ for
  $N\geq 3$, generally, only the parity of the real Chern number in Eq.~(2) is
  stable with trivial bands being added.}
\bibitem[{Pau()}]{Pauli-Matrices}
\bibinfo{note}{The Pauli matrices $\sigma_j$ appear in different expressions do
  not necessarily operate in the same space.}
\end{thebibliography}

\widetext
\clearpage


\section*{Supplementary material}

# Supplemental Materials: "Boundary criticality of $PT$-invariant topology and second-order nodal-line semimetals"

## CLIFFORD ALGEBRAS

For the Clifford algebras $C^{2n}$, there are $2n$ generators, represented by $2^n \times 2^n$ Dirac gamma matrices $\gamma^a$ with $a = 1, 2, ..., 2n$, satisfying

$$\{\gamma^a, \gamma^b\} = 2\delta^{ab}. \tag{S1}$$

Obviously, there is the $(2n+1)$th gamma matrix anti-commuting with all $\gamma^a$ with $a = 1, ..., 2n$, which is written as

$$\gamma^{2n+1} = \pm i^n \gamma^1 \gamma^2 \cdots \gamma^{2n}. \tag{S2}$$

Note that all gamma matrices are hermitian, $(\gamma^a)^\dagger = \gamma^a$, and $(\gamma^a)^2 = 1$, with $a = 1, ..., 2n+1$. In this paper, we choose $4 \times 4$ gamma matrices for $Cl^4$ as

$$\gamma^1 = \tau_0 \otimes \sigma_3, \gamma^2 = \tau_0 \otimes \sigma_1, \gamma^3 = \tau_2 \otimes \sigma_2, \gamma^4 = \tau_1 \otimes \sigma_2, \gamma^5 = \tau_3 \otimes \sigma_2, \tag{S3}$$

where $\gamma^{1,2,3}$ are pure real while $\gamma^{4,5}$ are pure imaginary.

## THE TOPOLOGICAL INVARIANT

The $PT$-symmetric Hamiltonian in our paper is

$$\mathcal{H}_0(\mathbf{k}) = \sin k_x \gamma^1 + \sin k_y \gamma^2 + (M - \cos k_x - \cos k_y)\gamma^3. \tag{S4}$$

Before calculating the topological invariants, we derive some equalities of the unitary transformation,

$$U = e^{-\frac{[\gamma^3, \gamma^5]}{2}\frac{\pi}{4}} = \cos\frac{\pi}{4} - \sin\frac{\pi}{4}\frac{[\gamma^3, \gamma^5]}{2}, \quad U^\dagger = \cos\frac{\pi}{4} + \sin\frac{\pi}{4}\frac{[\gamma^3, \gamma^5]}{2}. \tag{S5}$$

Obviously, $U, U^\dagger$ commute with $\gamma^1, \gamma^2, \gamma^4$, $U\gamma^i U^\dagger = \gamma^i, i = 1, 2, 4$, and

$$U\gamma^3 U^\dagger = \gamma^5, \quad U\gamma^5 U^\dagger = -\gamma^3. \tag{S6}$$

Thus we can obtain the block diagonalized Hamiltonian, $U\mathcal{H}_0(\mathbf{k})U^\dagger = \begin{pmatrix} \mathcal{H}_+(\mathbf{k}) & 0 \\ 0 & \mathcal{H}_-(\mathbf{k}) \end{pmatrix}$, where

$$\mathcal{H}_\pm(\mathbf{k}) = \sin k_x \sigma_3 + \sin k_y \sigma_1 \pm (M - \cos k_x - \cos k_y)\sigma_2 \tag{S7}$$

are two-bands Chern-insulators. Then we calculate their topological invariants. Eq. (S7) can be written as

$$\mathcal{H}_\pm(\mathbf{k}) = \mathbf{g}(\mathbf{k}) \cdot \boldsymbol{\sigma} \tag{S8}$$

with $\mathbf{g}(\mathbf{k}) = (\sin k_x, \sin k_y, \pm(\lambda - \cos k_x - \cos k_y))$, $\boldsymbol{\sigma} = (\sigma_3, \sigma_1, \sigma_2)$. The topological invariant (Chern number or winding number) can be calculated by

$$\nu = \frac{1}{24\pi^2} \int d^3\mathbf{k} \epsilon^{\mu\nu\rho} \operatorname{tr}\left(G\partial_\mu G^{-1} G\partial_\nu G^{-1} G\partial_\rho G^{-1}\right), \tag{S9}$$

where the imaginary-frequency Green's function is given as $G^{-1} = i\omega - \mathbf{g}(\mathbf{k}) \cdot \boldsymbol{\sigma}$. By using $\operatorname{tr}(\sigma_i \sigma_j \sigma_k) = 2i\epsilon_{ijk}$, we have

$$\nu = \frac{1}{4\pi^2}\epsilon^{ij}\epsilon_{nml} \int d\omega dk_x dk_y \frac{1}{(\omega^2 + |\mathbf{g}|^2)^2} g^n \partial_i g^m \partial_j g^l. \tag{S10}$$

Using the integral $\int_{-\infty}^\infty d\omega \frac{1}{(\omega^2+|\mathbf{g}|^2)^2} = \frac{\pi}{2|\mathbf{g}|^3}$, we can obtain

$$\nu = \frac{1}{4\pi} \int dk_x dk_y \frac{1}{|\mathbf{g}|^3} \mathbf{g} \cdot (\partial_{k_x}\mathbf{g} \times \partial_{k_y}\mathbf{g}) = \frac{1}{4\pi} \int dk_x dk_y\ \hat{\mathbf{g}} \cdot (\partial_{k_x}\hat{\mathbf{g}} \times \partial_{k_y}\hat{\mathbf{g}}), \tag{S11}$$

where $\hat{\mathbf{g}} = \frac{\mathbf{g}}{|\mathbf{g}|}$ is a unit vector. The Eq. (S11) is just the winding number from $S^2$ to $S^2$. Then the topological invariants of the Chern insulators can be computed,

$$\nu(\mathcal{H}_\pm(\mathbf{k})) = \begin{cases} \pm 1, & M \in (-2, 0), \\ \mp 1, & M \in (0, 2), \\ 0, & \text{otherwise.} \end{cases} \quad (S12)$$

The $PT$-symmetric Hamiltonian (S4) can be considered as a direct sum of two Chern insulators with the opposite Chern number when $-2 < m < 0$ and $0 < m < 2$. $\mathcal{H}_+(\mathbf{k})$ and $\mathcal{H}_-(\mathbf{k})$ are related to each other by $\hat{P}\hat{T} = \hat{\mathcal{K}}$ symmetry. This system is called a Stiefel-Whitney insulator with nontrivial $\mathbb{Z}_2$ topological charge [1]. The corresponding $\mathbb{Z}_2$ topological invariant can be calculated by

$$\nu_R = -\frac{1}{4\pi} \int_{BZ} \text{tr}(I\mathcal{F}_R) \mod 2, \quad (S13)$$

where $\mathcal{F}_R$ is the curvature for the real Berry bundle derived from the connection $\mathcal{A}_{\alpha\beta}^R = \langle \alpha, \mathbf{k} | d | \beta, \mathbf{k} \rangle$ with the real eigenstates $|\alpha, \mathbf{k}\rangle$ satisfying $|\alpha, \mathbf{k}\rangle = \hat{P}\hat{T}|\alpha, \mathbf{k}\rangle$. And $I$ is the generator of the $SO(2)$ group [2]. By this formula, the $\mathbb{Z}_2$ topological invariant of the Hamiltonian (S4) can be computed and equals to 1.

## THE BOUNDARY EFFECTIVE THEORY AND PROJECTORS

We now take the bottom boundary as an example to derive the boundary effective theory of the $PT$-symmetric Hamiltonian (S4). For the open boundary perpendicular to the $y$-direction, we apply the inverse Fourier transformation for $k_y$ to get the first Hamiltonian in real space as

$$\mathcal{H}_0(k_x) = \sin k_x \gamma^1 + \frac{1}{2i}(S_y - S_y^\dagger)\gamma^2 + \left[m - \cos k_x - \frac{1}{2}(S_y + S_y^\dagger)\right]\gamma^3, \quad (S14)$$

where $S_y$ is the translation operator along the $y$-direction as

$$S_y|i\rangle = |i+1\rangle, \quad S_y^\dagger|i\rangle = |i-1\rangle \quad (S15)$$

with integer $i$ labeling the lattice site of the $y$-direction. Accordingly, the matrices can be explicitly written as

$$S_y = \begin{pmatrix} \ddots & \vdots & \vdots & \vdots & \vdots & \cdot^{\cdot^{\cdot}} \\ \ddots & 0 & 0 & 0 & 0 & \cdots \\ \cdots & 1 & 0 & 0 & 0 & \cdots \\ \cdots & 0 & 1 & 0 & 0 & \cdots \\ \cdots & 0 & 0 & 1 & 0 & \cdots \\ \cdot^{\cdot^{\cdot}} & \vdots & \vdots & \vdots & \vdots & \ddots \end{pmatrix}, \quad S_y^\dagger = \begin{pmatrix} \ddots & \vdots & \vdots & \vdots & \vdots & \cdot^{\cdot^{\cdot}} \\ \ddots & 0 & 1 & 0 & 0 & \cdots \\ \cdots & 0 & 0 & 1 & 0 & \cdots \\ \cdots & 0 & 0 & 0 & 1 & \cdots \\ \cdots & 0 & 0 & 0 & 0 & \cdots \\ \cdot^{\cdot^{\cdot}} & \vdots & \vdots & \vdots & \vdots & \ddots \end{pmatrix}.$$

For the open boundary with non-negative part of the $y$-axis, we have $\widehat{S}_y^\dagger|0\rangle = 0$. More explicitly, the semi-infinite translation operators are now written as

$$\widehat{S}_y = \begin{pmatrix} 0 & 0 & 0 & 0 & \cdots \\ 1 & 0 & 0 & 0 & \cdots \\ 0 & 1 & 0 & 0 & \cdots \\ 0 & 0 & 1 & 0 & \cdots \\ \vdots & \vdots & \vdots & \vdots & \ddots \end{pmatrix}, \quad \widehat{S}_y^\dagger = \begin{pmatrix} 0 & 1 & 0 & 0 & \cdots \\ 0 & 0 & 1 & 0 & \cdots \\ 0 & 0 & 0 & 1 & \cdots \\ 0 & 0 & 0 & 0 & \cdots \\ \vdots & \vdots & \vdots & \vdots & \ddots \end{pmatrix}.$$

By taking the ansatz $|\psi_\mathbf{k}\rangle = \sum_{i=0}^\infty \lambda^i |i\rangle \otimes |\xi\rangle$, with $|\lambda| < 1$, we solve the Schrödinger equation of Eq. (S14). In the bulk with $i \geq 1$, we have

$$\left[\sin k_x \gamma^1 + \frac{1}{2i}(\lambda^{-1} - \lambda)\gamma^2 + \left(m - \cos k_x - \frac{1}{2}(\lambda^{-1} + \lambda)\right)\gamma^3\right]|\xi\rangle = \mathcal{E}|\xi\rangle. \quad (S16)$$

In the boundary with $i = 0$, we have

$$\left[\sin k_x \gamma^1 - \frac{1}{2i}\lambda\gamma^2 + \left(m - \cos k_x - \frac{1}{2}\lambda\right)\gamma^3\right]|\xi\rangle = \mathcal{E}|\xi\rangle \tag{S17}$$

The difference of Eqs. (S16) and (S17) gives

$$i\gamma^2\gamma^3|\xi\rangle = |\xi\rangle, \tag{S18}$$

which implies the boundary state is the eigenstate of $i\gamma^2\gamma^3$ with the eigenvalue as 1. Then, the projector for this state can be constructed as

$$\Pi^B = \frac{1}{2}(1 + i\gamma^2\gamma^3). \tag{S19}$$

Applying the projector to Eq. (S17), we have

$$\sin k_x \gamma^1 |\xi\rangle = \mathcal{E}|\xi\rangle, \tag{S20}$$

The difference of Eqs. (S20) and (S17), together with Eq. (S18), gives

$$\lambda = m - \cos k_x. \tag{S21}$$

The effective Hamiltonian for the boundary state is just

$$\mathcal{H}^B_{eff}(\mathbf{k}) = \Pi^B \mathcal{H}_0(\mathbf{k})\Pi^B = \sin k_x \gamma^1 \Pi^B. \tag{S22}$$

Similarly, we can obtain the upper projector and effective Hamiltonian,

$$\Pi^U = \frac{1}{2}(1 - i\gamma^2\gamma^3), \quad \mathcal{H}^U_{eff} = \sin k_x \gamma^1 \Pi^U, \tag{S23}$$

the left and right projectors and the corresponding effective Hamiltonians,

$$\Pi^L = \frac{1}{2}(1 + i\gamma^1\gamma^3), \quad \mathcal{H}^L_{eff} = \sin k_x \gamma^2 \Pi^L, \quad \Pi^R = \frac{1}{2}(1 - i\gamma^1\gamma^3), \quad \mathcal{H}^R_{eff} = \sin k_x \gamma^2 \Pi^R. \tag{S24}$$

Obviously, the projectors act on the $\gamma$'s in Eq. (S3) as

$$\Pi^{U/B}\gamma^{2,3}\Pi^{U/B} = 0, \quad \Pi^{R/L}\gamma^{1,3}\Pi^{U/B} = 0,$$
$$\Pi^{U/B}\gamma^{1,4,5}\Pi^{U/B} = \gamma^{1,4,5}\Pi^{U/B}, \quad \Pi^{R/L}\gamma^{2,4,5}\Pi^{R/L} = \gamma^{2,4,5}\Pi^{R/L}, \tag{S25}$$

which have been used for the derivation of the boundary Hamiltonians

$$\mathcal{H}^B_{eff} = \gamma^1(k_x + im_1\gamma^4 + im_3\gamma^5)\Pi^B, \quad \mathcal{H}^L_{eff} = \gamma^2(k_y + im_2\gamma^4 + im_4\gamma^5)\Pi^L, \tag{S26}$$

$PT$ symmetry connects the projectors of the opposite surfaces:

$$\hat{P}\hat{T}\Pi^B = \Pi^U \hat{P}\hat{T}, \quad \hat{P}\hat{T}\Pi^R = \Pi^L \hat{P}\hat{T}, \tag{S27}$$

which implies the effective Hamiltonians of opposite surfaces are connected by $PT$ symmetry as

$$\hat{P}\hat{T}H^B_{eff} = H^U_{eff}\hat{P}\hat{T}, \quad \hat{P}\hat{T}H^R_{eff} = H^L_{eff}\hat{P}\hat{T}. \tag{S28}$$

## NUMERICAL SIMULATION OF TIGHT-BANDING MODELS

### 2D Tight-Binding Model

The numerical simulations of 2D tight-binding model with perturbations are shown in Fig. 1(a,b,c,d). The Hamiltonian is

$$\mathcal{H}^{2D}(\mathbf{k}) = \sin k_x \gamma^1 + \sin k_y \gamma^2 + (1 - \cos k_x - \cos k_y)\gamma^3 + i\gamma^1(m_1\gamma^4 + m_3\gamma^5) + i\gamma^2(m_2\gamma^4 + m_4\gamma^5). \tag{S29}$$

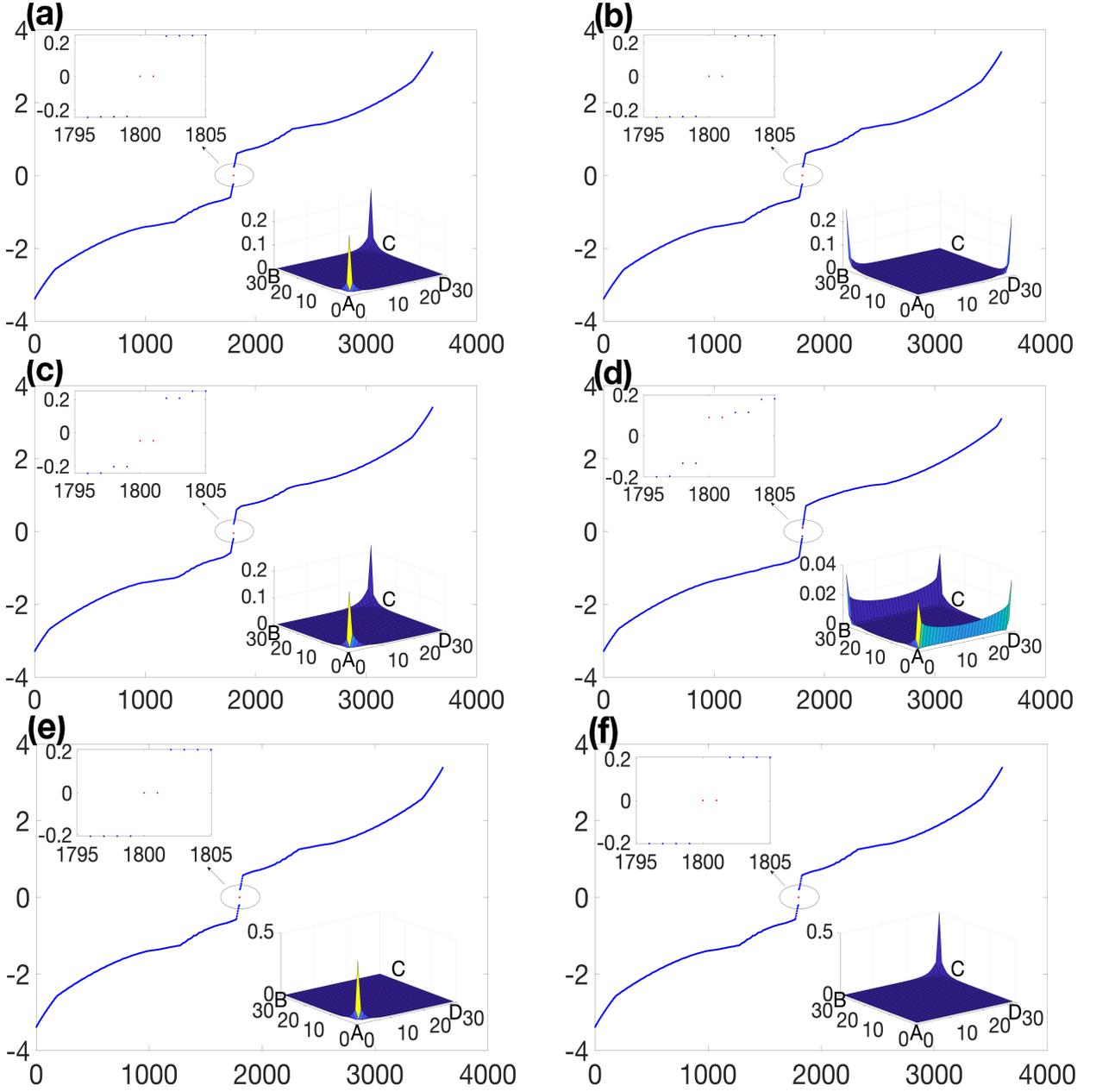

Figure 1. We plot the eigen-energy of $30 \times 30$ lattice sites with open boundary condition. The probabilities of the states inside the gap are shown by the right-down insets of the subfigures. (a) $(m_1, m_3) = (m_2, m_4) = (0.2, 0.2)$ for $k = 1$. (b) $(m_1, m_3) = -(m_2, m_4) = (0.2, 0.2)$ for $k = -1$. Corner states are moved to the corners of $B$ and $D$ as discussed in the main text. (c) $(m_1, m_3) = (0.2, 0.2)$ and $(m_2, m_4) = (0.1, 0.2)$. There exists corner states of nonzero energy. (d) $(m_1, m_3) = (0.2, 0)$ and $(m_2, m_4) = (0, -0.1)$. There exists no corner states. (e), (f). The real part of the eigen-energy of the Hamiltonian with additional non-Hermitian perturbation $i\epsilon\gamma^4$ (or $i\epsilon\gamma^5$). The probabilities of two real zero-energy states (the two red dots in left-up insets of the subfigures) are shown in (e) and (f) espectively. The parameters $m_i$ with $i = 1, 2, 3, 4$ are same as condition (a) and $\epsilon = 0.2$.

## 3D Tight-Binding Model

The numerical simulations of 3D tight-binding model with perturbations are shown in Fig. 2(a,b). The corresponding Hamiltonian is

$$\mathcal{H}^{3D}(\mathbf{k}) = \sin k_x \gamma^1 + \sin k_y \gamma^2 + (M - \cos k_x - \cos k_y - \cos k_z)\gamma^3 + i\gamma^1(m_1\gamma^4 + m_3\gamma^5) + i\gamma^2(m_2\gamma^4 + m_4\gamma^5). \quad (S30)$$

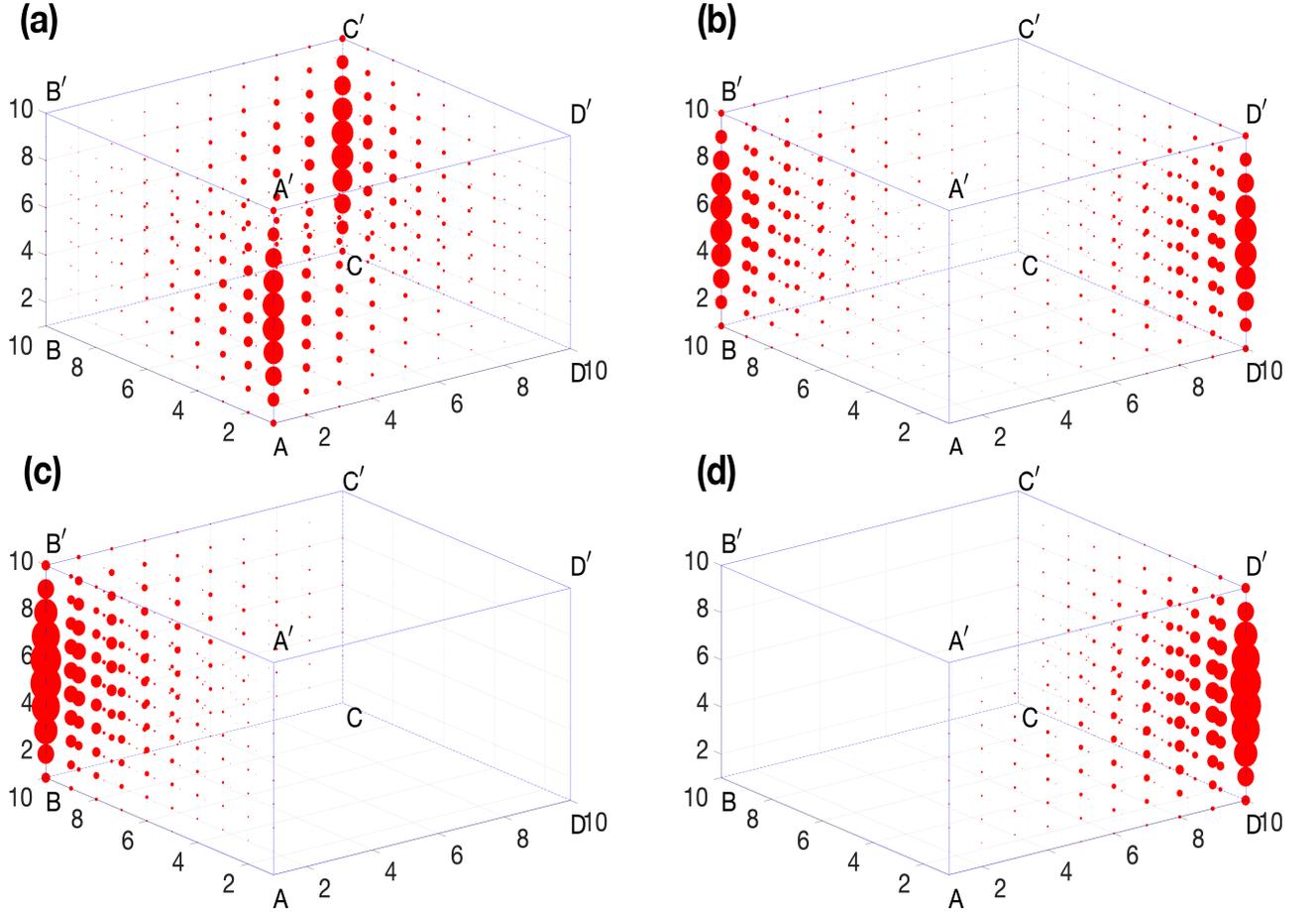

Figure 2. We plot the probabilities of the hinge states on the $10 \times 10 \times 10$ lattice sites. $M = 2$. (a) $(m_1, m_3) = (m_2, m_4) = (0.2, 0.2)$ for $k = 1$. The hinge states are located at the edges of $AA'$ and $CC'$. (b) $(m_1, m_3) = (0.2, 0.2)$ and $(m_2, m_4) = (-0.2, -0.2)$ for $k = -1$. The hinge states are located at the edges of $BB'$ and $DD'$. (c), (d). The probabilities of the hinge states of the 3D Dirac semimental with additional non-Hermitian perturbation $i\epsilon\gamma^4$ (or $i\epsilon\gamma^5$). $(m_1, m_3) = -(m_2, m_4) = (0.2, 0.2)$, for $k = -1$, and $\epsilon = 0.2$. The hinge states are the middle two real part zero-mode states. They are related to each other by $PT$ symmetry, since their corresponding imaginary-part energies conjugate to each other.

**The Tight-Binding Model with Non-Hermitian Perturbations**

As a further disscussion, we consider the 2D tight-binding model (S29) with an additional non-Hermitian perturbation,

$$\mathcal{H}_{NH}(\mathbf{k}) = \mathcal{H}^{2D}(\mathbf{k}) + i\epsilon\gamma^4, \tag{S31}$$

where the non-Hermitian term $i\epsilon\gamma^4$(or $i\epsilon\gamma^5$) breaks the implicit chiral symmetry $\Lambda_1$ but preserves the $PT$ symmetry. In this case, the zero-energy states can be localized at only one corner. In contrast to the recent research on second-order topological phases in non-Hermitian systems [3], the two real part zero-energy modes in our model (S31) are localized at corner $A$ and $C$ (or corner $B$ and $D$) respectively. The numerical simulation results are shown in Fig. 1(e,f). The two real part zero-modes are related by $PT$ symmetry, since their corresponding imaginary-part energies conjugate to each other.

On the other hand, the hinge states in 3D Dirac semimental stacked by non-Hermitian 2D Hamiltonian (S31) and trivial insulators are similar to Fig. 2, except that each has one hinge state, and the two hinge states are related by $PT$ symmetry, since their corresponding imaginary-part energies conjugate to each other. The numerical simulations are shown in Fig. 2(c,d).



\end{document}